\documentclass[aps,pre,twocolumn]{revtex4}
\usepackage{epsfig}
\begin{document}
\def\I.#1{\it #1}
\def\B.#1{{\bf #1}}
\def\C.#1{{\cal  #1}}
\title{Scaling Exponents in Anisotropic Hydrodynamic Turbulence}
   \author { Victor S. L'vov$^*$, Itamar Procaccia$^*$ and Vasil Tiberkevich$^*\dag$}
   \affiliation{$^*$Department of~~Chemical Physics, The Weizmann Institute of
   Science, Rehovot 76100, Israel\\
$^\dag$Radiophysical Faculty, National Taras Shevchenko 
University of Kiev, Kiev, Ukraine}
%
%
\begin{abstract}
In anisotropic turbulence the correlation functions are decomposed in
the irreducible representations of the SO(3) symmetry group (with
different ``angular momenta" $\ell$). For different
values of $\ell$ the second order correlation function
is characterized by different scaling exponents $\zeta_2(\ell)$. In this
paper we compute these scaling exponents in a Direct Interaction Approximation
(DIA). By linearizing the DIA equations in small anisotropy we set up
a linear operator and find its zero-modes in the inertial interval
of scales. Thus the scaling exponents in each $\ell$-sector follow from
solvability condition, and are 
not determined by dimensional analysis. The main result of our
calculation is that the scaling exponents
$\zeta_2(\ell)$ form a strictly increasing spectrum at least until $\ell=6$, guaranteeing
that the  effects of anisotropy decay as power laws when the scale of observation
diminishes. The results of our calculations are compared to available
experiments and simulations.

\end{abstract}
   \maketitle
\section{Introduction}

All realistic turbulent flows are maintained by anisotropic (and
inhomogeneous) forcing.
Thus the principal conceptual model of turbulence, i.e. 
``homogeneous isotropic turbulence",
exists only in theory. Testing theoretical predictions that are derived
on the basis of such a model in experimental flows (or in simulations)
that are patently anisotropic can sometime lead to premature or erroneous
conclusions about important issues like the universality of scaling
exponents and other fundamental issues in the theory of turbulence.
The justification for disregarding the effects of anisotropy was the old
conjecture that in the limit of very high Reynolds numbers and very small scales, local
isotropy may be restored by the non-linear transfer mechanism that cascades energy from
large to small scales. In the last decade there had been a number of observation that
claimed the opposite
\cite{95PS,96Pum,98GW}. On the whole, these observations were based on measuring objects that
``should vanish" for isotropic flows, and observing their behavior as
a function of Reynolds number (Re) or scale. Thus for example objects made of the
normal derivative of the downstream velocity components were examined:
 \begin{equation}
{\cal S}_{2k+1} \equiv \frac{\langle \left(\partial u_x/\partial y\right)^{2k+1}\rangle}
{\langle \left(\partial u_x/\partial y\right)^2\rangle ^{k+1/2}} \ . \label{skew}
\end{equation}
The pointed brackets denote ensemble average, $u_x$ is the streamwise component of the Eulerian
velocity field $\B.u(\B.r)$,  and $y$ is in the spanwise direction. Since such
objects vanish in isotropic systems, their increase as a function of Re were interpreted as a lack of
restoration of local isotropy. The problem with such measures is that objects of this type are also
sensitive to the phenomenon of intermittency, and also perfectly isotropic objects like
 \begin{equation}
{\cal K}_{2k+1} \equiv \frac{\langle \left(\partial u_x/\partial x\right)^{2k+1}\rangle}
{\langle \left(\partial u_x/\partial x\right)^2\rangle ^{k+1/2}} \ ,
\end{equation}
increase with Reynolds number. It is thus unclear what is the more important
source of the observation that the objects (\ref{skew}) do not vanish even
when the Reynolds number is increased.

Until rather
recently it was not obvious how to assess the anisotropic effects in a clear fashion,
separating the contributions of the isotropic sector from the rest. Starting with
\cite{ALP}, it was proposed that one can do so usefully by finding systematically 
the projections of the measured correlation or structure functions on the
irreducible representations of the SO(3) group of all rotations. This approach
was found useful in analyzing experimental results \cite{98ADKLPS,00KLPS,00KS}
and numerical simulations \cite{99ABMP,02BDLT}. In the context of passive
scalar and passive scalar advection it gave rise to a number of exact
results \cite{00ABP,00ALPP,01AP}. In its simplest form the projection is
applied to the $p$th order structure functions (with $\hat\B.R\equiv \B.R/R$, $R\equiv |\B.R|$):
\begin{equation}
S_p(\B.R) \equiv \langle \left[\B.u(\B.x+\B.R)-\B.u(\B.x))\cdot\hat\B.R\right]^p\rangle \ .
\end{equation}
Such objects admit a relatively simple SO(3) decomposition since they are scalar objects.
We can thus span them by the usual spherical harmonics:
\begin{equation}
S_p(\B.R) =\sum_{\ell=0}^\infty\sum_{m=-\ell}^\ell S_p^{\ell m}(R) Y_{\ell m}(\hat\B.R) \ .
\end{equation}
In this equation we have used the indices $\ell m$ to label,
respectively, the total angular momentum and its projection along a reference
axis, say $\hat \B.z$. We are interested in particular in the scaling properties
of the amplitudes $S_p^{\ell m}(R)$,
\begin{equation}
S_p^{\ell m}(R) \sim A_{\ell m} R^{\zeta_p(\ell)} \ .
\end{equation}
In the case of exactly soluble models \cite{00ABP,00ALPP,01AP} it was found that
the scaling exponents $\zeta_p(\ell)$ form a strictly increasing spectrum
as a function of $\ell$. In such cases it becomes completely clear that for $R\to 0$,
which in the limit Re $\to \infty$ can still be in the inertial range, the
higher order $\ell$ contributions disappear in favor of the isotropic
contribution alone. Thus if one can demonstrate the existence of a strictly
increasing spectrum of exponents also in the context of the Navier-Stokes 
dynamics, one would establish that local isotropy is restored at the small
scales when Re $\to \infty$ (i.e. when the viscous cut-off goes to zero). The numerical values
of the scaling exponents will determine the rate (in scales) at which isotropy
is restored. The aim of this paper is to present such a calculation.

A calculation of the second order anisotropic exponents $\zeta_2(\ell)$, based on the
Navier-Stokes equations, was attempted before in \cite{01GHL}. The analysis
there concentrated on the {\em forced} solutions for the second order
structure function $S_2(\B.R)$, and concluded with two sets of 
dimensional predictions.
The first, assuming that the anisotropic forcing is analytic, reads
\begin{equation}
\zeta_2(\ell) = \ell+2/3 \ , \quad\text{forced solution, analytic} \ .\label{forcedana}
\end{equation} 
The second forced solution was computed for non analytic forcing, resulting with 
\begin{equation}
\zeta_2(\ell) = 4/3 \ , \quad\text{forced solution, nonanalytic} \ .\label{forcednot}
\end{equation} 
Another, more phenomenological approach, was presented in \cite{02BDLT}, generalizing
an earlier argument by Lumley \cite{67Lum}. In this approach
one does not balance the energy transfer against the forcing, but rather invokes
the existence of a shear $s_{ik}\equiv \partial u_i/\partial r_k$ 
as the main reason for the anisotropy.   Performing dimensional analysis
in which the shear is added to $\bar\epsilon$, the mean energy flux per unit time and mass,
one ends up with the prediction
\begin{equation}
\zeta_2(\ell) =\frac{2+\ell}{3} \ , \quad \text{dimensional, shear dominated.} \label{shear}
\end{equation}
We note that for $\ell=2$ the predictions (\ref{forcednot}) and 
(\ref{shear}) coincide;  all three predictions disagree
for $\ell>2$.

These predictions do not agree with the result of the only vector
model with pressure that had been solved exactly, i.e. the  
``linear pressure model" \cite{01AP}. This model captures some of the aspects of the
pressure term in Navier-Stokes turbulence, while being linear and therefore much
simpler problem. The non linearity of the Navier-Stokes equation is
replaced by an advecting field $\B.w(\B.x,t)$ and an advected field
$\B.v(\B.x,t)$. The advecting field $\B.w(\B.x, t)$ is taken with known
dynamics and statistics.  Both fields are assumed incompressible. The
equation of motion for the vector field $v^\alpha(\B.x, t)$ is:
\begin{eqnarray}
    \partial_t v^\alpha + w^\mu \partial_\mu v^\alpha +
        \partial^\alpha p - \kappa \partial^2 v^\alpha &=& f^\alpha \ ,
        \label{eq:v-p} \\
    \partial_\alpha v^\alpha = 0 \ , \quad
    \partial_\alpha w^\alpha &=& 0 \ . \nonumber
\end{eqnarray}
In this equation, $\B.f(\B.r,t)$ is a divergence free forcing term and
$\kappa$ the viscosity. The domain of the system is taken to be infinite.
Following Kraichnan's model for passive scalar \cite{68Kra}, the
advecting field $\B.w(\B.r,t)$ is chosen to be a white-noise Gaussian process with a
correlation function which is given by:
\begin{eqnarray}
    &&\delta(t'-t)D^{\alpha\beta}(\B.R) \equiv 
        \left< w^\alpha(\B.r + \B.R, t') w^\beta(\B.r, t) \right> \ ,\\
     &&K^{\alpha\beta}(\B.R) \equiv 
         D^{\alpha\beta}(\B.R) - D^{\alpha\beta}(\B.0) \nonumber\\
      &&= DR^\xi\left[(\xi+2)\delta^{\alpha\beta} - 
               \xi\frac{R^\alpha R^\beta}{R^2}\right] \ . \label{defxi}
\end{eqnarray}
The forcing $\B.f(\B.x,t)$ is also taken to be a Gaussian white noise
process. Its correlation function is
\begin{equation}
  F^{\alpha\beta} (\B.R/L)\delta(t-t') \equiv 
   \langle f^\alpha(\B.r+\B.R,t) f^\beta(\B.r,t')\rangle \ .
\end{equation}
The forcing is responsible for injecting energy and anisotropy to the
system at an outer scale $L$. We choose the tensor function
$F^{\alpha\beta}(\B.x)$ to be analytic in $\B.x$, anisotropic,
and vanishing rapidly for $|\B.x|\gg 1$.

To compare with the predictions (\ref{forcedana}) and (\ref{shear}) we
should take $\xi=4/3$ in Eq. (\ref{defxi}). For this value of $\xi$ the 
result of \cite{01AP} are the exponents $\zeta_2(0)=2/3$, 
$\zeta_2(2)=1.25226$, $\zeta_2(4)=2.01922$, $\zeta_2(6)=4.04843$,
$\zeta_2(8)=6.06860$ and $\zeta_2(10)=8.08337$, in rather
sharp disagreement with the predictions (\ref{forcedana})--(\ref{shear}).
We will see that the calculation presented below for the Navier-Stokes case
comes up with results in close agreement with those of the 
linear pressure model. We thus will present a strong belief that
the dimensional predictions (\ref{forcedana})--(\ref{shear}) fail to capture
the correct results for the Navier-Stokes case.

In our approach we start from the Navier-Stokes equations, and
write down an approximate equation satisfied by the second order
correlation function, in the Direct Interaction Approximation (DIA). This
equation is nonlinear. For a weakly anisotropic system we can linearize the
equation, to define a linear operator over the space of the anisotropic
components of the second order correlation function. The solution is then
a combination of forced solutions and ``zero modes" which are eigenfunctions
of eigenvalue zero of the linear operator. The exponents of the forced solutions are
identical to (\ref{forcedana}), but the exponents of the zero modes are
smaller, and therefore leading with respect to former. 
The exponents (\ref{forcednot}) are not physical, and are not observed
in experiments or simulations. The exponents of the zero modes
are close to (\ref{shear}) for $\ell=2$ and 4, but begin to deviate strongly
for $\ell=6$, falling very close to the predictions of the linear pressure model. We will
argue that again the exponents of the zero modes are those that are observed in simulations.

The structure of the paper is as follows: in Sect. \ref{modeleqs}
we set up the DIA equations for the second order structure function, and
linearize them in weak anisotropy. We present the symmetry properties of
the resulting operator, to simplify as much as possible the SO(3) decomposition
which is presented in Sect. \ref{SO3}. The actual calculation of the 
scaling exponents is detailed in Sect. \ref{calculation}. Finally in Sect. \ref{conclusions}
we present concluding remarks.

\section{Model Equations for Weak Anisotropy in the DIA approximation}
\label{modeleqs}
\subsection{DIA equations}
It is customary to discuss the DIA equations in $\B.k,t$ representation.
The Fourier transform of the velocity field $\B.u(\B.r,t)$ is defined by
\begin{equation}\label{ukt}
{\B.u}({\B.k},t)\equiv
\int d{\B.r}\, \exp[-i({\B.r}\cdot{\B.k}    )] {\B.u}({\B.r},t) \ .
\end{equation}
The Navier-Stokes equations for an incompressible fluid then read
\begin{eqnarray}\nonumber
 &&  \Big[ {\partial \over \partial t}+\nu k^2\Big]u^\alpha  (\B.k,t) = \frac{i}
 {2}\Gamma ^{\alpha \beta \gamma }(\B.k)\int {d^3 q d^3 p\over (2\pi)^3}
  \\    \label{NSE:kt}
&\times &\delta (\B.k+\B.q+\B.p){u^*}^\beta (\B.q,t) {u^*}^\gamma  (\B.p,t) \ .
\end{eqnarray}
The interaction amplitude $\Gamma ^{\alpha \beta \gamma }(\B.k)$ is defined
by
\begin{equation}
\Gamma ^{\alpha \beta \gamma }(\B.k) =-\left[P^{\alpha\gamma}(\B.k) k^\beta
+P^{\alpha\beta}(\B.k)k^\gamma \right] \ ,
\end{equation}
with the transverse projection operator $P^{\alpha\beta}$ defined as
\begin{equation}
P^{\alpha\beta} \equiv \delta^{\alpha\beta} -\frac{k^\alpha k^\beta}{k^2} \ .
\end{equation}
The statistical object that is the concern of this paper is the second
order (tensor) correlation function $\B.F(\B.k,t)$,
\begin{equation}
(2\pi)^3 F^{\alpha\beta}(\B.k,t)\delta (\B.k-\B.q) \equiv \langle u^\alpha
(\B.k,t){u^*}^\beta(\B.q,t)\rangle \ .
\end{equation}
In stationary conditions
this object is time independent. Our aim is to find its $k$-dependence,
especially in the anisotropic sectors.

It is well known that there is no close-form theory for the second order
simultaneous correlation function. We therefore need to resort to standard
approximations that lead to model equations. An approach that is by now
time-honored is Kraichnan's DIA, which 
leads to the approximate equation of motion
\begin{equation}
\frac{\partial F^{\alpha\beta}(\B.k,t)}{2\partial t} = I^{\alpha\beta}
(\B.k,t)-\nu k^2 F^{\alpha\beta}(\B.k,t)  \ ,
\end{equation}
where
\begin{equation}
I^{\alpha\beta}(\B.k) = \int\frac {d^3q d^3p}{(2\pi)^3}
\delta(\B.k+\B.p+\B.q)
\Phi^{\alpha\beta} (\B.k,\B.q,\B.p) \ .  \label{Integral}
\end{equation}
In this equation
\begin{equation}
\Phi^{\alpha\beta} (\B.k,\B.q,\B.p) =\frac{1}{2}[\Psi^{\alpha\beta} (\B.k,\B.q,\B.p)
+\Psi^{\beta\alpha} (\B.k,\B.q,\B.p) ] \ , \
\end{equation}
and 
\begin{eqnarray}
&&\Psi^{\alpha\beta} (\B.k,\B.q,\B.p) = \Theta(\B.k,\B.q,\B.p)
\Gamma^{\alpha\gamma\delta}(\B.k)\nonumber\\
&&\times[\Gamma^{\delta\beta'\gamma'}(\B.q)
 F^{\gamma\gamma'}(\B.p) F^{\beta'\beta}(\B.k) \nonumber\\&&+
\Gamma^{\gamma\beta'\delta'}(\B.p)
 F^{\delta\delta'}(\B.q) F^{\beta'\beta}(\B.k) \nonumber\\
&&+ \Gamma^{\beta\delta'\gamma'}(\B.k)
 F^{\delta\delta'}(\B.q) F^{\gamma\gamma'}(\B.p)] \ . \label{Phi}
\end{eqnarray}
In stationary conditions and for $k$ in the inertial range we need
to solve the integral equation $I^{\alpha\beta}(\B.k) = 0$.

The process leading to these equations is long; one starts with the
Dyson-Wyld perturbation theory, and truncates (without justification)
at the first loop order. In addition one asserts that the time dependence
of the response function and the correlation functions are the same.
Finally one assumes that the time correlation functions decay in time
in a prescribed manner. This is the origin of the ``triad interaction 
time" $\Theta(\B.k,\B.q,\B.p)$.
If one assumes that all the correlation functions involved decay exponentially
(i.e. like $\exp(-\gamma_\B.k|t|)$, then
\begin{equation}
\Theta(\B.k,\B.q,\B.p) =\frac{1}{\gamma_\B.k+\gamma_\B.q+\gamma_\B.p} \ . \label{expdecay}
\end{equation}
For Gaussian decay, i.e. like $\exp[-(\gamma_\B.k t)^2/2]$,
\begin{equation}
\Theta(\B.k,\B.q,\B.p) =\frac{1}{\sqrt{\gamma^2_\B.k+\gamma^2_\B.q+\gamma^2_\B.p}} \ .
\label{gaussdecay}
\end{equation}
All these approximations are uncontrolled. Nevertheless DIA is known to
give roughly correct estimates of scaling exponents and even of coefficients.
For the case at hand, where we are interested in scaling exponents that were
never computed from first principle, it certainly pays to examine what this
approach has to predict.

Eq. (\ref{Integral}) poses a nonlinear
integral equation which is closed once we model $\gamma_\B.k$. One may 
use the estimate $\gamma_\B.k\sim k U_k$ where
$U_k$ is the typical velocity amplitude on the inverse scale of $k$,
which is evaluated as $U^2_k\sim k^3 F^{\alpha\alpha}(\B.k)$.
\begin{equation}
\gamma_\B.k =  C_\gamma k^{5/2}\sqrt {F^{\alpha\alpha}(\B.k)} \ .
\label{gamma}
\end{equation}
In isotropic turbulence Eqs. (\ref{Integral}) and (\ref{gamma}) have an exact
solution with K41 scaling exponents, 
\begin{eqnarray}
F^{\alpha\beta}_0(\B.k) &=& P^{\alpha\beta}(\B.k) F(k) \ , \nonumber\\
F(k) &=& C\bar\epsilon^{2/3} k^{-11/3}\ , \quad
\gamma_k= \tilde C_\gamma \bar\epsilon^{1/3} k^{2/3} \ . \label{j0}
\end{eqnarray}
Note that the scaling exponents in $\B.k$-representation have
a $d$-dependent difference from their numerical value in $\B.r$-
representation. In 3-dimensions $\zeta_2\to\xi_2= \zeta_2+3$, and the
exponent 2/3 turns to 11/3 in Eq.(\ref{j0}).

For weak anisotropic turbulence Eq.(\ref{Integral}) will pose a {\em linear}
problem for the anisotropic components which depends on this isotropic solution.
\subsection{DIA with Weak Anisotropy} 

In weakly anisotropic turbulence we consider a small anisotropic correction 
$f^{\alpha\beta} (\B.k)$ to the 
fundamental isotropic background
\begin{equation}
F^{\alpha\beta}(\B.k)=F^{\alpha\beta}_0(\B.k) + 
f^{\alpha\beta} (\B.k) \ . \label{smallf}
\end{equation}
The first term vanishes with the solution (\ref{j0}). Linearizing our
integral equation with respect to the anisotropic correction we read
\begin{eqnarray}
&&I^{\alpha\beta}(\B.k)\!=\!\! \int\!\!\frac {d^3q d^3p}{(2\pi)^3}
\delta(\B.k+\B.p+\B.q)
[S^{\alpha\beta\gamma\delta} (\B.k,\B.q,\B.p) f^{\gamma\delta}(\B.k) \nonumber\\
&&+2T^{\alpha\beta\gamma\delta} (\B.k,\B.q,\B.p) f^{\gamma\delta}(\B.q)]=0 \
, \nonumber\\ &&S^{\alpha\beta\gamma\delta} (\B.k,\B.q,\B.p)\equiv
\frac{\delta\Phi^{\alpha\beta} (\B.k,\B.q,\B.p)}{\delta
F^{\gamma\delta}(\B.k)}\ ,\nonumber\\ &&T^{\alpha\beta\gamma\delta}
(\B.k,\B.q,\B.p)\equiv \frac{\delta\Phi^{\alpha\beta} (\B.k,\B.q,\B.p)}{\delta
F^{\gamma\delta}(\B.q)} \ . \label{allj}
\end{eqnarray}
We reiterate that the functional derivatives in Eq.(\ref{allj}) are calculated in
the isotropic ensemble. In computing these derivatives we should account 
also for the implicit dependence of  $\Theta(\B.k,\B.q,\B.p)$ on the correlation
function through Eq. (\ref{gamma}).
We can rewrite Eq. (\ref{allj}) in a way that brings out
explicitly the linear integral operator $\hat L$,
\begin{equation}
\hat L |\B.f\rangle\equiv \int\!\!\frac {d^3q}{(2\pi)^3}
{\cal L}^{\alpha\beta\gamma\delta}(\B.k,\B.q) f^{\gamma\delta}(\B.q) = 0
\ , \label{opereq}
\end{equation}
where the kernel of the operator is
\begin{eqnarray}
{\cal L}^{\alpha\beta\gamma\delta}(\B.k,\B.q)&\equiv&
\delta(\B.k-\B.q)\int\frac {d^3p}{(2\pi)^3}S^{\alpha\beta\gamma\delta} 
(\B.k,\B.p,-\B.k-\B.p)
\nonumber\\ &&+2T^{\alpha\beta\gamma\delta} (\B.k,\B.q,-\B.k-\B.q) \ . \label{oper}
\end{eqnarray}

\subsection{Symmetry properties of the linear operator}

The first observation to make is that the linear operator is invariant
under all rotations. Accordingly we can block diagonalize it by expanding
the anisotropic perturbation
in the irreducible representation of the SO(3) symmetry group.
These have principal indices $\ell$ with an integer $\ell$ going from
0 to $\infty$. The zeroth component is the isotropic sector. 
Correspondingly our integral equation takes the form
\begin{equation}
I^{\alpha\beta}(\B.k) = I^{\alpha\beta}_0(\B.k)+\sum_{\ell=1}^\infty 
I^{\alpha\beta}_\ell(\B.k) =0
\ . \label{intj}
\end{equation}
The block diagonalization implies that each $\ell$-block provides an independent
set of equations (for every value of $\B.k$):
\begin{equation}
I^{\alpha\beta}_\ell(\B.k) =0 \ .
\end{equation}
The first term of (\ref{intj}) vanishes with the solution (\ref{j0}). 
For all higher values of $\ell$ we need to solve the corresponding
equation
\begin{equation}
\hat L |\,\B.f_\ell\rangle = 0
\ . \label{operell}
\end{equation}
We can block diagonalize further by exploiting additional symmetries
of the linear operator. In all our discussion we assume that our turbulent flow has zero
helicity. Correspondingly all the correlation functions are invariant
under the inversion of $\B.k$:
\begin{equation}
F_0^{\alpha\beta}(\B.k) =F_0^{\alpha\beta}(-\B.k)\ , \quad 
f^{\alpha\beta}_\ell (\B.k)=f^{\alpha\beta}_\ell (-\B.k) \ , \label{symmF1}
\end{equation}
Consequently there are no odd $\ell$ components, and
we can write
\begin{equation}
f^{\alpha\beta} (\B.k)= \sum_{j=2,4,...}^\infty
f^{\alpha\beta}_\ell (\B.k) \ . \label{fullF}
\end{equation}
We also note that in general $\B.u(-\B.k)=\B.u^*(\B.k)$. Accordingly, the correlation
functions are real. From this fact and the definition it follows that the correlation
functions are symmetric to index permutation, 
\begin{equation}
F^{\alpha\beta}_0(\B.k) =F^{\beta\alpha}_0(\B.k)\ , \quad 
f^{\alpha\beta}_\ell (\B.k)=f^{\beta\alpha}_\ell (\B.k) \ . \label{symmF2}
\end{equation}
As a result our linear operator is invariant to permuting the first ($\alpha,\beta$)
and separately the second ($\gamma,\delta$) pairs of indices. In addition,
the operator is symmetric to $\B.k\to -\B.k$ together with $\B.q\to -\B.q$. This
follows from the symmetry (\ref{symmF1}) and from the appearance of products of
two interaction amplitudes (which are antisymmetric under the inversion
of all wave-vectors by themselves).

Finally, our kernel is a homogeneous function of the wavevectors, meaning that in 
every block we can expand in terms of basis functions that have a definite
scaling behavior, being proportional to $k^{-\xi}$.

\section{SO(3) decomposition}
\label{SO3}
As a result of the symmetry properties the operator $\hat L$ is block 
diagonalized by tensors that have the following properties:
\begin{itemize}
\item They belong to a definite sector $(\ell, m)$ of the SO($3$) group.
\item They have a definite scaling behavior, i.e., are proportional to
  $k^{-\zeta_2}$ with some scaling exponent $\zeta_2$.
\item They are either symmetric or antisymmetric under permutations of
  indices.
\item They are either even or odd in $\B.k$.
\end{itemize}
In \cite{00ABP} we discuss these types of tensors in detail. Here we only quote
the final results. In every sector $(\ell, m)$ of the rotation group with
$\ell > 1$, one can find 9 independent tensors $X^{\alpha\beta}(\B.k)$ that
scale like $k^{-\xi_2(\ell)}$. They are given by $k^{-\xi_2(\ell)} \tilde B_{j,\ell
m}^{\alpha\beta}(\hat{\B.k})$, where the index $j$ runs from 1 to 9, enumerating the
different spherical tensors. The unit vector $\hat\B.k\equiv \B.k/k$. These nine 
tensors can be further subdivided into four subsets:
\begin{itemize}
\item \textbf{Subset I} of 4 symmetric tensors with $(-)^\ell$ parity.  
\item \textbf{Subset II} of 2 symmetric tensors with $(-)^{\ell+1}$ parity.
\item \textbf{Subset III} of 2 antisymmetric tensors with $(-)^{\ell+1}$ parity.
\item \textbf{Subset IV} of 1 antisymmetric tensor with $(-)^\ell$ parity.
\end{itemize}
Due to the diagonalization of $\hat L$ by these subsets, the equation for
the zero modes foliates, and we can compute the zero modes in each subset
separately. In this paper, we choose to focus on subset I, which has
the richest structure. The four tensors in this subset are given by
\begin{eqnarray}
  \tilde B_{1,\ell m}^{\alpha\beta}(\hat{\B.k}) &=& 
        k^{-\ell-2}k^\alpha k^\beta \phi_{\ell m}(\B.k) \nonumber , \\
 \tilde B_{2,\ell m}^{\alpha\beta}(\hat{\B.k}) &=& 
        k^{-\ell}[k^\alpha \partial^\beta + k^\beta\partial^\alpha] 
          \phi_{\ell m}(\B.k) \ , \nonumber \\
 \tilde B_{3,\ell m}^{\alpha\beta}(\hat{\B.k}) &=& 
        k^{-\ell}\delta^{\alpha\beta} \phi_{\ell m}(\B.k) \ , \nonumber \\
 \tilde B_{4,\ell m}^{\alpha\beta}(\hat{\B.k}) &=& 
        k^{-\ell+2}\partial^\alpha \partial^\beta \phi_{\ell m}(\B.k) \ ,
       \label{basis}
\end{eqnarray}
where $\phi_{\ell m}(\B.k)$ are the standard spherical harmonics 
We expect the calculation of the other subsets to be easier.

The last property to employ is the incompressibility of our target
function $f^{\alpha\beta} (\B.k)$. Examining the basis (\ref{basis})
we note that we can find two linear combinations that are
transverse to $\B.k$ and two linear combinations that are
longitudinal in $\B.k$. We need only the former, which have
the form
\begin{eqnarray}
 B_{1,\ell m}^{\alpha\beta}(\hat{\B.k}) &=& 
        k^{-\ell}P^{\alpha\beta}(\B.k) \phi_{\ell m}(\B.k) \nonumber , \\
 B_{2,\ell m}^{\alpha\beta}(\hat{\B.k}) &=& 
        k^{-\ell}[k^2 \partial^\alpha \partial^\beta -(\ell-1)( k^\beta \partial^\alpha
+k^\alpha\partial^\beta)\nonumber\\&&+\ell(\ell-1) \delta^{\alpha\beta} ]
          \phi_{\ell m}(\B.k) \ . \label{basis}
\end{eqnarray}
Using this basis we can now expand our target function as
\begin{equation}
\label{expand}
f^{\alpha\beta}_\ell (\B.k) = k^{-\xi_2(\ell)} 
   \Big[ c_1 B_{1,\ell m}^{\alpha\beta}(\hat{\B.k}) + 
         c_2 B_{2,\ell m}^{\alpha\beta}(\hat{\B.k})  
    \Big] \ . 
\end{equation}
\section{Calculation of the scaling exponents}
\label{calculation}
Substituting Eq.(\ref{expand}) into Eq.(\ref{operell}) we find
\begin{equation}
\hat L q^{-\zeta_2(\ell)}|\B.B_{1,\ell m}\rangle c_1+\hat L q^{-\xi_2(\ell)}|\B.B_{2,\ell
m}\rangle c_2 =0 \ . \label{LonB}
\end{equation}
Projecting this equation on  the two function of the basis (\ref{basis}) we obtain
a matrix
\begin{eqnarray}
&&L_{i,j}(\ell, \xi_2(\ell)) \equiv \langle \B.B_{i,\ell m}|\hat L q^{-\xi_2(\ell)}|
\B.B_{j,\ell m}\rangle\label{mateq}\\
&&= \int\!\!\frac {d^3q}{(2\pi)^3}\,d\hat \B.k\,B^{\alpha\beta}_{i,\ell
m}(\hat\B.k)
{\cal L}^{\alpha\beta\gamma\delta}(\B.k,\B.q)q^{-\xi_2(\ell)}B^{\gamma\delta}_{j,\ell
m}(\hat\B.q) \nonumber\ .
\end{eqnarray}
Here we have full integration with respect to $\B.q$, but only 
angular integration with respect to $\B.k$. Thus the matrix depends on $k$ as
a power, but we are not interested in this dependence since we demand the solvability
condition
\begin{equation}
\det L_{i,j}(\ell, \xi_2(\ell)) = 0 \ . \label{solvability}
\end{equation}
It is important to stress that in spite of the explicit
$m$ dependence of the basis functions, the matrix obtained in this way
has no $m$ dependence. In the calculation below we can therefore put,
without loss of generality, $m=0$. This is like having cylindrical symmetry
with a symmetry axis in the direction of the unit vector $\hat\B. n$.
In this case we can write the matrix $\B. B_{i, \ell}(\hat\B. k)$ (in the
vector space $\alpha,\, \beta= x,\,y,\,z$) as   
\begin{equation}\label{B-as-operator}   
B^{\alpha\beta}_{i, \ell}(\hat\B. k) = k^{-\ell} 
\hat {\C. B}^{\alpha\beta}_{i, \ell,\B. k} (k^{\ell} P_{\ell}
(\hat\B. k \cdot \hat\B. n))\,,
\end{equation}   
where $\hat {\C. B}^{\alpha\beta}_{i, \ell,\B. k}$ are matrix
operators, acting on wave vector $\B. k$:
\begin{eqnarray}\label{def-B-operator}   
\hat {\C. B}^{\alpha\beta}_{1, \ell,\B. k}
& \equiv  & \delta^{\alpha\beta} -   
\frac{k^\alpha k^\beta}{k^2}, \\  \nonumber 
\hat {\C. B}^{\alpha\beta}_{2, \ell,\B. k}
& \equiv  & \frac{ k^2\, \partial^2}{ \partial k^\alpha \partial k^\beta}
- (\ell - 1) \Big ( \frac{   k^\alpha \partial}{\partial k^\beta}
 +\frac{   k^\beta \partial}{\partial k^\alpha } 
- \ell \, \delta^{\alpha\beta}\Big ) \, ,   
\end{eqnarray}   
and $P_\ell(x)$ denote $\ell$-th order Legendre polynomials.
\subsection{Angular Averaging}
To proceed, we perform the angular   
averaging in Eq.~(\ref{mateq}) (i.e., integration over 
all directions of $\hat\B. k$) analytically.  In order
to do this we note that Eq.~(\ref{mateq}), after substituting
Eq. (\ref{B-as-operator}), is invariant to the simultaneous
rotation of the vectors $\B. k$, $\B. q$, and $\hat\B. n$.  This means that
after integrating over $\B. q$, Eq.~(\ref{mateq}) must have the form
\begin{eqnarray} 
L_{i,j}(\ell, \xi_2) &=&  \int d\hat\B. k M_{i, j, \ell,   
\xi_2}(k, \hat\B. k \cdot \hat\B. n) \nonumber\\ 
 &=& \int d\hat\B. n M_{i,   
j, \ell, \xi_2}(k, \hat\B. k \cdot \hat\B. n) \ ,
\end{eqnarray}
where $\B. M$ is an appropriately defined matrix.
Accordingly, we 
can change the integration over $\hat\B. k$ in favor of integrating over   
$\hat\B. n$.  Thus instead of having the direction $\hat\B.n$ fixed
and all the other vector rotating, we will now choose the direction of
$\B.k$ fixed, and rotate the other vectors. Note also that operator
$\hat\B. L$ does not depend    on $\hat\B. n$, and only the matrices $\B. B_{i, \ell}$
are     averaged upon.  Thus Eq.~(\ref{mateq}) can be written as    
\begin{equation}\label{}   
L_{i,j}(\ell, \xi_2) = \int \frac{d\B. q}{(2 \pi)^3}   
L^{\alpha\beta\gamma\delta}(\B. k, \B. q) q^{-\xi_2}   
\Lambda^{\alpha\beta\gamma\delta}_{i j, \ell}(\hat\B. k,   
\hat\B. q),    
\end{equation}   
where    
\begin{eqnarray}\label{Lambda}   &&
\Lambda^{\alpha\beta\gamma\delta}_{i j, \ell}(\hat\B. k, \hat\B. q) \equiv \int
d\hat\B. n \, B^{\alpha\beta}_{i, \ell}(\hat\B. k) B^{\gamma\delta}_{j,
\ell}(\hat\B. q) \\ \nonumber 
& = & 4\pi (2\ell+1) k^{-\ell} q^{-\ell} \hat{\C. B}^{\alpha\beta}_{i,
\ell,\B. k} \hat{\C. B}^{\gamma\delta}_{j, \ell,\B. q}\left[ k^\ell
q^\ell P_\ell (\hat\B. k \cdot \hat\B. q) \right]\ .
\end{eqnarray}   
Here we used the definition (\ref{B-as-operator})   
and the following property of the Legendre polynomials   
\begin{equation}\label{}   
\int d\hat\B. n P_\ell(\hat\B. n \cdot \hat\B. k) P_\ell(\hat\B. n \cdot \hat\B. q) =   
4\pi(2\ell + 1) P_\ell(\hat\B. k \cdot \hat\B. q).   
\end{equation}  

Now, using Eq.~(\ref{def-B-operator}) we can write the $\Lambda$-matrices
explicitly:
\begin{widetext} \onecolumngrid
\begin{equation}\label{}   
\Lambda^{\alpha\beta\gamma\delta}_{ij,\ell}(\hat\B. k, \hat\B. q) =   
P^{\alpha\alpha'}_{\B. k}P^{\beta\beta'}_{\B. k}
P^{\gamma\gamma'}_{\B. q}P^{\delta\delta'}_{\B. q}   
\tilde{\Lambda}^{\alpha'\beta'\gamma'\delta'}_{ij,\ell}(\hat\B. k,   
\hat\B. q),   
\end{equation}   
where   
\begin{eqnarray}\label{}   
\tilde{\Lambda}^{\alpha\beta\gamma\delta}_{11,\ell}(\hat\B. k,   
\hat\B. q)  & = &   
\delta^{\alpha\beta}\delta^{\gamma\delta}P_\ell(\hat\B. k \cdot  
\hat\B. q)   \,,    \\ 
\tilde{\Lambda}^{\alpha\beta\gamma\delta}_{12,\ell}(\hat\B. k,   
\hat\B. q)  & = & \tilde{\Lambda}^{\gamma\delta\alpha\beta}_{21,\ell}
(\hat\B. q,   
\hat\B. k)   = \delta^{\alpha\beta}\delta^{\gamma\delta}   
\left[\ell^2 P_\ell(\hat\B. k \cdot \hat\B. q)  -  (\hat\B. k \cdot \hat\B. q)   
  P'_\ell(\hat\B. k \cdot \hat\B. q) \right]  +    
\delta^{\alpha\beta}\hat{k}^\gamma\hat{k}^\delta   
P''_\ell(\hat\B. k \cdot \hat\B. q)    \,,
\\   
\tilde{\Lambda}^{\alpha\beta\gamma\delta}_{22,\ell}(\hat\B. k,   
\hat\B. q)   
& = & \delta^{\alpha\beta}\delta^{\gamma\delta} \left[  
\ell^4 P_\ell(\hat\B. k \cdot \hat\B. q) - (2\ell^2-1)(\hat\B. k \cdot  
\hat\B. q)P'_\ell(\hat\B. k \cdot \hat\B. q)\right.  + \left. (\hat\B. k \cdot \hat\B. q)^2   
P''_\ell(\hat\B. k \cdot \hat\B. q)   
\right]  + (\delta^{\alpha\gamma}\delta^{\beta\delta}    
+\delta^{\alpha\delta}\delta^{\beta\gamma})P''_\ell(\hat\B. k \cdot  
\hat\B. q)   
\nonumber\\   
&& + 
\hat q^\alpha \hat q^\beta \hat k^\gamma \hat k^\delta 
P^{(IV)}_\ell(\hat\B. k \cdot \hat\B. q)   
 +  \left( \delta^{\alpha\beta}\hat k^\gamma \hat k^\delta   
+\hat q^\alpha \hat q^\beta \delta^{\gamma\delta}\right)   
\left[(\ell^2-1)P''_\ell(\hat\B. k \cdot \hat\B. q)    
  -(\hat\B. k \cdot \hat\B. q)P'''_\ell(\hat\B. k \cdot \hat\B. q)\right]   
\nonumber\\   
&& +  \left(   
\hat q^\alpha \delta^{\beta\gamma} \hat k^\delta +    
\hat q^\alpha \delta^{\beta\delta} \hat k^\gamma +   
\hat q^\beta \delta^{\alpha\gamma} \hat k^\delta +    
\hat q^\beta \delta^{\alpha\delta} \hat k^\gamma \right)  
P'''_\ell(\hat\B. k \cdot \hat\B. q)  \ .  
\end{eqnarray}  

\end{widetext} \twocolumngrid
\subsection{Transform to 2-dimensional Integral} 
\label{2D}
Examining Eq.~(\ref{mateq}) we recall that the matrix ${\cal L}^{\alpha\beta\gamma\delta}(\B.k,\B.q)$
contains an integration over $\B.p$, cf. Eq. (\ref{oper}). This
integration is relatively trivial because of the existence
of the $\delta$-function.
We can integrate over
$\B. p$ simply expressing $\B. p$ as $\B. p = -\B. k -\B. q$. Next
we integrate over $\B. q$ in spherical coordinates. Since we fixed the
direction of $\B.k$, we can choose it, without loss of generality,
in the direction of the $\hat\B. z$ axis. Then,
\begin{equation}\label{}   
L_{i,j} (\ell,\xi_2) = \int_0^{+\infty} q^2 dq    
\int_0^{\pi} \sin \Theta d\Theta \int_0^{2\pi} d\phi 
\tilde{L}_{i,j}(\B. k,\B. q).   
\end{equation}   
It can be shown that    
$\tilde{L}_{i,j}(\B. k,\B. q) = \tilde{L}_{i,j}(k,q,\cos   
\Theta)$, i.e. does not depend on angle $\phi$.   
So, we obtain a 2-dimensional integral   
\begin{equation}\label{finint}   
L_{i,j} (\ell,\xi_2) = \int_0^{+\infty} q^2 dq    
\int_{-1}^{+1} da \tilde{L}_{i,j}(k,q,a),   
\end{equation}   
where $a = \cos \Theta$.   
   
One more remark: the kernel in Eq.~(\ref{mateq}) is symmetric with respect to
permuting the vectors $\B. q$ and $\B. p$. This means that we can actually
integrate not over all $\B. q$-space, but only over half-space,
namely, when $q<p=\sqrt{k^2+2a k q +q^2}$. This not only
decreases the calculation time, but also allows us not to integrate near the
point $p \approx 0$, where the kernel is in general 
singular.

\subsection{Window of Locality}
\label{locality}
In performing the integration numerically we need to worry about
the convergence of the integrand. Convergence is guaranteed only
within a given interval of the scaling exponent $\zeta_2(\ell)$ which
is referred to as the ``window of locality".   
To find the window of locality one should expand the kernel in Eq.~(\ref{finint})
for both small and large $q$ and investigate its behavior at these
regions. It is a straightforward (but cumbersome) procedure, and we show
explicit results of such an expansion only near $q \approx 0$ for $\ell = 4$
and the `exponential' decay model (\ref{expdecay}). Also we choose here $k=1$,
exploiting the homogeneity of all our operators in $k$. The
equations satisfied by $\tilde{L}_{i,j}(k=1,q,a)$ are:
\begin{widetext} \onecolumngrid
\begin{eqnarray}\label{xxx}   
q^2 \tilde{L}_{1,1} & = &   
-\frac{5}{216}a (1 - a^2 ) ( 3 - 30a^2 + 35a^4)q^{3 - \xi_2}   
 +\frac{5}{288}a(1 - a^2)( 3 - 30\,a^2 + 35a^4)q^{11/3 - \xi_2}   
 \\  \nonumber   
  && - \frac{1}{648}(1-a^2)(3 - 52a^2) ( 3 - 30a^2 + 35a^4)q^{4-\xi_2}    
   -\frac{5}{432}a(1 - a^2)( 3 - 30a^2 + 35a^4 )q^{13/3 -   
     \xi_2}   
   + O(q^{14/3-\xi_2})   \,, 
\nonumber \\  \nonumber \\   
q^2 \tilde{L}_{1,2} & = &   
\frac{5}{18}a( 1 - 6a^2 + 5a^4)q^{3 - \xi_2}   
-\frac{5}{144}a( 1 + 9a^2 - 45a^4 + 35a^6)q^{11/3 - \xi_2}   
 \nonumber \\   
 &&+\frac{1}{108}(1 - a^2)( 51 - 494a^2 + 835a^4)q^{4 - \xi_2}   
 + \frac{5}{144}a( -1 + 21a^2 - 55a^4 + 35a^6)q^{13/3 -   
   \xi_2}   
   + O(q^{14/3-\xi_2})   
 \nonumber \\   
\nonumber \\   
q^2 \tilde{L}_{2,1} & = &\frac{5}{18}a( -3 + 30a^2 - 55a^4 +28a^6)q^{3 - \xi_2}   
 - \frac{5}{24}a( -3 + 30a^2 - 55a^4 + 28a^6)q^{11/3 - \xi_2}   
 \nonumber \\   
 && + \frac{1}{108}(1 - a^2)
( -9 + 426a^2 - 3105a^4 + 3080a^6)q^{4 - \xi_2}   
 + \frac{5}{36}a( -3 + 30a^2 - 55a^4 + 28a^6)q^{13/3 -   
   \xi_2}  + O(q^{14/3-\xi_2})   
 \nonumber \\   
\nonumber \\   
q^2 \tilde{L}_{2,2} & = &   
-\frac{5}{6}a( 13 - 30a^2 + 17a^4)q^{3 - \xi_2}   
 - \frac{5}{3}a(1 - a^2)^2 ( -4 + 7a^2)q^{11/3 - \xi_2}   
 \nonumber \\   
 && - \frac{1}{18}(1 - a^2)( -96 + 89a^2 + 155a^4)q^{4 - \xi_2}   
 -\frac{5}{24}a(1 - a^2)( 19 - 71a^2 + 56a^4)q^{13/3 - \xi_2}   
   + O(q^{14/3-\xi_2})   
\end{eqnarray}   
\end{widetext} \twocolumngrid 
After integration over $a=\cos \Theta$ we obtain   
\begin{eqnarray}\label{yyy}   
\int_{-1}^1 da \, q^2 \tilde{L}_{1,1} & = & -   
\frac{832}{25515}q^{4-\xi_2}+ O(q^{14/3-\xi_2})   \,,
\\   
\int_{-1}^1 da \, q^2 \tilde{L}_{1,2} & = &    
\frac{832}{2835}q^{4-\xi_2}    + O(q^{14/3-\xi_2})\,,   
\nonumber \\   
\int_{-1}^1 da \, q^2 \tilde{L}_{2,1} & = &     
-\frac{4544}{8505}q^{4-\xi_2}    + O(q^{14/3-\xi_2})  \,, 
\nonumber \\   
\int_{-1}^1 da \, q^2 \tilde{L}_{2,2} &=&    
\frac{4544}{945}q^{4-\xi_2}   + O(q^{14/3-\xi_2})\ .
   \nonumber 
\end{eqnarray}   
It is clear that integrals have IR divergence if $\xi_2 > \xi^* = 5$.
   
This result may seem surprising, since
the original kernel in Eq.~(\ref{mateq}) has terms that depend on $q$ as
$q^{-\xi_2}$. Each of these terms begins to diverge if $\xi_2 >3$. There
is, however, a cancellation of the leading terms, resulting in an increase in the
IR limit of the window of locality, up to the $\xi_2 =4$. The subleading
terms turn out to be antisymmetric in $a$, always vanishing after
the angular integration. Thus the actual limit of the window of locality
is as computed above.
   
The situation is even more complicated for the next anisotropic sector
$\ell = 6$. The next subleading term (sub-subleading), which gives the
main IR contribution in the case $\ell = 4$, also vanishes after
integration. This is due to the fact that the matrix elements $\tilde{L}_{i,j}$
contain Legendre polynomials $P_\ell(a)$ as multipliers; these are
orthogonal to all $a^n$, $n<\ell$, and the highest order of
$a$ in the term $q^{m-\xi_2} a^n$, Eq.~(\ref{xxx}),  can not be greater than
$m+2$. So, one can conclude that the integrals converge in IR regime up
to $\xi_2 < \xi^*(\ell) = \ell+1$, for $\ell\ge 4$.  For $\ell \le 4$
we have $\xi^* = 5$.
   
The UV boundary of the window of locality also moves if $\ell$
increased, for the same reasons.
\subsection{Integrals near the IR Edge of the  Window of Locality:
approximate calculation of the exponents}   
\label{nearedge}   
It is clear from Eq.~(\ref{yyy}) that each integral near the   
critical point $\xi_2 \approx \xi^* = 5$    
(but $\xi_2 < \xi^*$!)   
has the form    
\begin{equation}\label{}   
L_{i,j}(\xi_2) = \frac{\alpha_{i,j}}{\xi^*-\xi_2} + \beta_{i,j}(\xi_2)
\,,
\end{equation}   
where $\alpha_{i,j}$ are given by the main coefficients in Eq.~(\ref{yyy}),
and $\beta_{i,j}(\xi_2)$ are regular functions near the point $\xi_2
\approx \xi^*$.
   
The main observation is that
\begin{equation}\label{}   
\det(\alpha_{i,j}) = \alpha_{1,1} \alpha_{2,2}- \alpha_{1,2}
\alpha_{2,1}=0 \,,
\end{equation}   
i.e. the determinant of the leading (divergent) parts of the integrals
vanishes. This occurs equally well for $\ell=4$ and $\ell=6$,
independently of decay model. Thus the full determinant can be written as
\begin{eqnarray}\label{detmod}   
\det(L_{i,j})& = &
\displaystyle{\frac{\alpha_{1,1}\beta_{2,2}+\alpha_{2,2}\beta_{1,1}
-\alpha_{1,2}\beta_{2,1}-\alpha_{2,1}\beta_{1,2}}{\xi^*-\xi_2}} 
\nonumber\\   
 && +(\beta_{1,1}\beta_{2,2}-\beta_{1,2}\beta_{2,1})\ .
\end{eqnarray}   
Thus the determinant diverges in general at $\xi_2=\xi^*$.   
For $\xi_2 \approx \xi^*$ the determinant   
is determined predominantly by the divergent term   
$\propto 1/(\xi^*-\xi_2)$. 

We can use this fact to estimate the scaling exponents:  
in zeroth approximation one can use Eq.~(\ref{detmod}) with 
$\beta_{i,j}(\xi_2)$  calculated exactly at the point $\xi_2=\xi^*$.   
The approximate value of scaling exponent is then
\begin{equation}\label{}   
\xi_2=\xi^* +
\displaystyle{\frac{\alpha_{1,1}\beta_{2,2}+\alpha_{2,2}\beta_{1,1}
-\alpha_{1,2}\beta_{2,1}-\alpha_{2,1}
\beta_{1,2}}{\beta_{1,1}\beta_{2,2}-\beta_{1,2}\beta_{2,1}}} \ .
\end{equation}   
This estimate is valid only as long as $1\gg\xi^*-\xi_2>0$. Actually,
the values of $\xi_2$ estimated this way for both $\ell=4$ and $\ell=6$
and the two decay models yield $\xi^*-\xi_2 \approx 0.01 - 0.02$, validating
the zeroth order approximation.  It is possible, however, to
calculate the determinant in this region exactly, as is done in the
next subsections.
   
\subsection{Calculating the Integrals near   
the IR Edge}   
\label{edge}   
Let us denote the integrands in Eq.~(\ref{mateq}) after the   
integration over $\cos \Theta$ as $J_{i,j}(q)$. Then we have   
\begin{equation}\label{}   
L_{i,j}=\int_0^{+\infty} J_{i,j}(q) dq   \ .
\end{equation}   
Let us also introduce   
\begin{eqnarray}\label{}   
I_{i,j}(q_0)&=&\int_{q_0}^{+\infty}J_{i,j}(q)dq\,, \\   
\delta I_{i,j}(q_0)&=& \int_0^{q_0}J_{i,j}(q)dq   \ .
\end{eqnarray}   
Then we have   
\begin{equation}\label{}   
L_{i,j}=I_{i,j}(q_0)+\delta I_{i,j}(q_0)   
\end{equation}   
for an arbitrary $q_0$.   
   
For $q_0 \ne 0$ $I_{i,j}(q_0)$ can be calculated numerically directly,
because there are no singularities for $q \ne 0$. (Note, and cf.
Subsect. \ref{2D}, we integrate over half $\B. q$-space, which does not include
the second singular point $p=\sqrt{k^2+2akq+q^2}=0$). On the other hand,
using Eq.~(\ref{yyy}) $\delta I_{i,j}(q_0)$ for sufficiently small $q_0$ and
$\xi_2 \approx \xi^*$ can be represented as
\begin{equation}\label{}   
\delta I_{i,j}(q_0) =   
\alpha_{i,j}\frac{q_0^{\xi^*-\xi_2}}{\xi^*-\xi_2} + O(q_0^{2/3})   
\end{equation}   
and one obtains the following formula for the integral $L_{i,j}$:   
\begin{equation}\label{Lmod}   
L_{i,j} \approx L_{i,j}(q_0) \equiv I_{i,j}(q_0) +   
\alpha_{i,j}\frac{q_0^{\xi^*-\xi_2}}{\xi^*-\xi_2}  \ . 
\end{equation}   
The test of validity of this formula is the independence   
of $L_{i,j}(q_0)$ of $q_0$.  
   
We have computed $L_{i,j}$ using Eq.~(\ref{Lmod})
with $q_0$ varying over a wide range. It turns out that $L_{i,j}(q_0)$ is practically
independent  of $q_0$ provided that $q_{\max}>q_0>q_{\min}$, where:
\begin{eqnarray}\label{}   
\ell = 4: &\,\,\,q_{\max} \approx 2\cdot10^{-3} &\,, \,\,\,   
q_{\min} \approx 5\cdot 10^{-8}\,; \nonumber \\   
\ell = 6: &\,\,\,q_{\max} \approx 2\cdot10^{-2} &\,, \,\,\,   
q_{\min} \approx 2\cdot 10^{-4}\ .\nonumber    
\end{eqnarray}   
``Practically independent of'' means that the integrals change in this
region of $q_0$ by an amount that is smaller than the minimum error of integration,
(and see next subsection for the estimate of this error).
   
For $q_0 > q_{\max}$ the simple approximation for $\delta
I_{i,j}(q_0)$ is not valid. For $q_0<q_{\min}$ the error of
integration starts to grow rapidly. This is connected with high-order
cancellations and finite machine precision. So, for $\ell=6$ we have a
4th-order cancellation, which means that any small error in the calculation
(or representation) of the leading term will increase by the factor
$q_0^{-4}$. The relative precision in presenting numbers on the machine
is about $10^{-16}$, so for $q_0 = q_{\min}(\ell=6)=2\cdot 10^{-4}$ we
have a principal (machine) relative error of about 0.1, and, of course,
the error of integration is increased here.
   
\subsection{Integration Method}   
   
To perform the integration over $q$ we used a standard Simpson integration
rule which gives errors of the order $f^{(4)}(x) h^4$ (here $f(x)$ -
integrand, $h$ - integration step).
   
Performing the integration over $a=\cos \Theta$ we used a 9-point closed
type Newton-Cotes integration formula with error of the order
$f^{(10)} h^{10}$. We have to use such high-order integration formula
because of high-order cancellation for $\ell=4$ and $\ell=6$. Simpler 
integration schemes amplify small relative 
errors in the integration of the leading terms (which should cancel after angular
integration) causing great absolute errors for small $q$.
   
The precision of integration was estimated by integral   
recalculation with smaller step $h$, and the error was   
determined as the maximal difference between the integrals,   
calculated with steps $h$, $h/2$ and $h/4$.   
   
The error in determining the scaling exponent was estimated as   
$$
\Delta \xi_2 =  \frac{ \Delta \det(L_{i,j})}{
\partial \det(L_{i,j})/\partial \xi_2}\,, 
$$ 
where $\Delta \det(L_{i,j})$ is the accuracy of determining  $\det(L_{i,j})$.
We estimated $ \Delta \xi_2 <0.002$  in all cases.
   
\section{Results and Concluding remarks}
\label{conclusions}
\begin{figure}
\vskip 0.5cm
\hskip -1cm
\epsfxsize=8.5cm \epsfbox{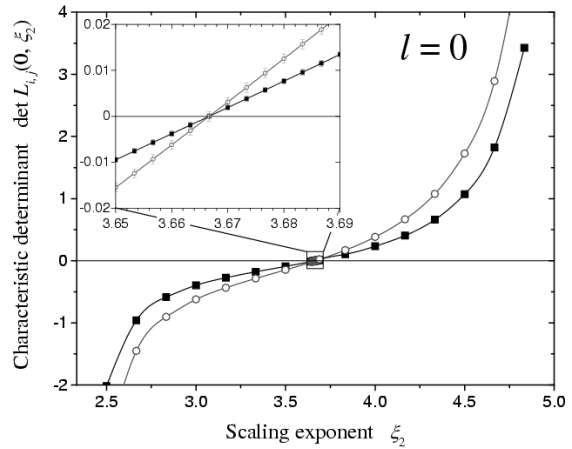}
\caption{determinant and zero crossing for the sector $\ell=0$.
The scaling exponent computed from the zero crossing is $\zeta_2(\ell=0)\approx
0.667$.}
\label{Fig.1}
\end{figure}

\begin{figure}
\vskip 0.5cm
\hskip -0.5cm
\epsfxsize=8cm \epsfbox{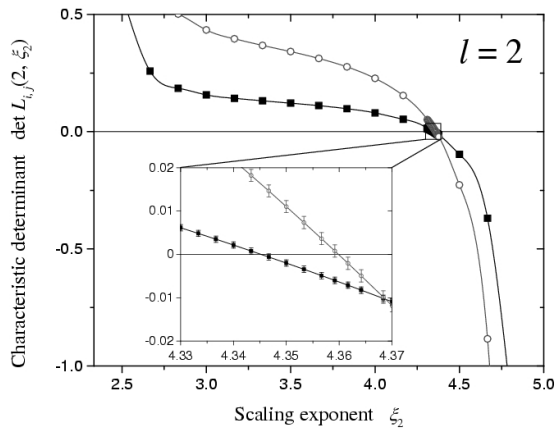}
\caption{determinant and zero crossing for the sector $\ell=2$.
The scaling exponent computed from the zero crossing is $\zeta_2(\ell=2)\approx
1.36-1.37$.}
\label{Fig.2}
\end{figure}

\begin{figure}
\vskip 0.5cm
\hskip -0.5cm
\epsfxsize=8cm \epsfbox{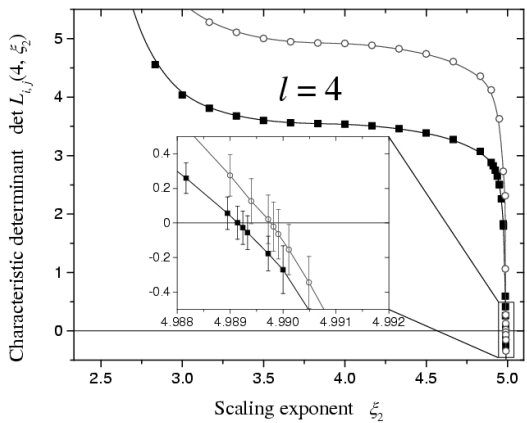}
\caption{determinant and zero crossing for the sector $\ell=4$.
The scaling exponent computed from the zero crossing is $\zeta_2(\ell=4)\approx
1.99$.}
\label{Fig.3}
\end{figure}

The determinants $\det[L_{i,j}(\ell,\xi_2)]$ were computed as a function of
the scaling exponents $\xi_2$ in every $\ell$-sector separately,
and the scaling exponent was determined from the zero crossing.
The procedure is exemplified in Fig. \ref{Fig.1} for the isotropic
sector $\ell=0$. We expect for this sector $\xi_2(0)=11/3$, in accordance
with $\zeta_2(0)=2/3$. Indeed, for both decay models, i.e the exponential decay (\ref{expdecay}), 
shown in dark
line, and the Gaussian decay (\ref{gaussdecay}) shown in light line, the 
zero crossing occurs at the 
same point, which in the inset can be read as 3.6667. 
For the higher $\ell$-sectors the agreement between the exponential
and gaussian models is not as perfect, indicating that our procedure
is not exact. In Fig. \ref{Fig.2} we present the determinant and
zero crossings for $\ell=2$. 
From the inset we can read the 
exponents $\xi_2(2)=4.351$ and 4.366 for the exponential and Gaussian
models respectively. This is in correspondence with $\zeta_2(2)=1.351$
and 1.366 respectively. These numbers are in excellent correspondence with
the experimental measurements reported in \cite{98ADKLPS,00KLPS}.
The results for $\ell=4$ are presented in Fig. \ref{Fig.3}. Here the zero
crossing, as seen in the inset, yields very close results for $\xi_2(4)$
between the exponential and Gaussian decay models, i.e. $\xi_2(4)\approx
4.99$. Note that this result is very close to the boundary of
locality as discussed in Subsect. \ref{locality}. Nevertheless the zero 
crossing is still easily resolved by the numerics, with
the prediction that $\zeta_2(4)\approx 1.99$. We note that
this number is well within the error bars of the simulational estimate of \cite{02BDLT}.

\begin{figure}
\vskip 0.5cm
\hskip -0.5cm
\epsfxsize=8cm \epsfbox{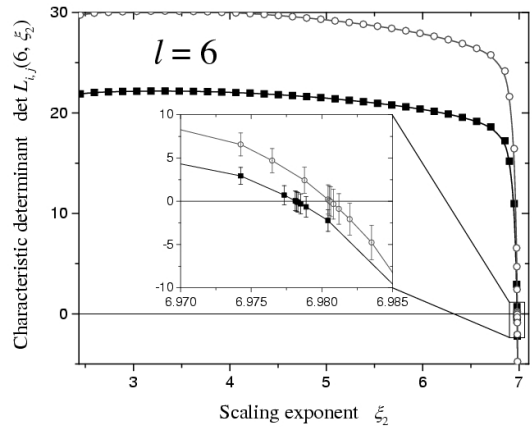}
\caption{determinant and zero crossing for the sector $\ell=6$.
The scaling exponent computed from the zero crossing is $\zeta_2(\ell=6)\approx
3.98$.}
\label{Fig.4}
\end{figure}

Similar results are
obtained for $\ell=6$, see Fig. \ref{Fig.4}. Also in this case 
exhibits zero crossing close to the boundary of locality, with
$\xi_2(6)\approx 6.98$. Again we find close correspondence between
the exponential and Gaussian models. In terms of $\zeta_2$ this means
$\zeta_2(6)\approx 3.98$. This number appears higher than the simulational
result of \cite{02BDLT}, which estimated $\zeta_2(6)\approx 3.2$. We note
however that for $\ell=6$ the log-log plots of \cite{02BDLT} scaled
over less than half a decade, and improved simulations may well result
in a substantial increase in the estimate.

Interestingly enough, the set of exponents $\zeta_2(\ell)$=2/3, 1.36, 1.99 and 3.98 for
$\ell=$0, 2, 4 and 6 respectively are in close agreement with the
numbers obtained for the linear pressure model, $\zeta_2(\ell)$=2/3, 1.25226, 2.01922, 4.04843,
for $\ell=0,2,4$ and 6 respectively. We reiterate at this point that the latter
set is exact for the linear pressure model, whereas the former set is
obtained within the DIA approximation. Nevertheless the close correspondence 
between the two leads us to propose that the actual exponents in the Navier-Stokes
case must be very close to these predictions. We thus propose that careful
experiments and simulations are likely to find the anisotropic exponents
$\zeta_2(\ell)\approx \ell-2$ for all $\ell>2$, with about 2/3 and 4/3 for
$\ell=0$ and 2 respectively. If so, the restoration of isotropy at large
Re and small scales should be quite clear, with high $\ell$ contributions
decaying very rapidly, and the $\ell=2$ contribution decaying with a
gap exponent of about 2/3. We do not expect a much more precise theoretical
evaluation of these exponents before the intermittency problem in the isotropic
sector is fully settled.

\acknowledgments  
This work has been supported in part by the European Commission
under a TMR grant, The Minerva Foundation, Munich, Germany, the German Israeli Foundation, 
the Israeli Science Foundation and the
Naftali and Anna Backenroth-Bronicki Fund for Research in
Chaos and Complexity.


\begin{thebibliography}{99}

\bibitem{95PS}
A. Pumir and B. Shraiman, Phys. Rev. Lett. {\bf 75}, 3114 (1995).

\bibitem{96Pum}
A. Pumir, Phys. Fluids {\bf 8}, 3112 (1996).

\bibitem{98GW}
S. Garg and Z. Warhaft, Phys. Fluids {\bf 10}, 662 (1998).

\bibitem{ALP}
I. Arad, V.S. L'vov and I. Procaccia, Phys. Rev. E {\bf 59}, 6753 (1999).


\bibitem{98ADKLPS}
I. Arad, B. Dhruva, S. Kurien, V.S. L'vov, I. Procaccia and K.R. Sreenivasan, Phys. Rev.
       Lett., {\bf 81}, 5330 (1998).

\bibitem{00KLPS}
S. Kurien, V. S. L'vov, I. Procaccia and K.R. Sreenivasan, 
       Phys. Rev. E{\bf 61}, 407 (2000). 

\bibitem{00KS}
S. Kurien and K. R., Sreenivasan, Phys. Rev. E{\bf 62}, 2206 (2000).

\bibitem{99ABMP}
I. Arad, L. Biferale, I. Mazzitelli and I. Procaccia, Phys. Rev. Lett.
{\bf 82}, 5040 (1999).

\bibitem{02BDLT}
L. Biferale, I. Daumont, A. Lanotte and F. Toschi, preprint (2002).

\bibitem{00ABP}
I. Arad, L. Biferale and I. Procaccia,  Phys.Rev.E, {\bf 61}, 2654 (2000).

\bibitem{00ALPP}
I. Arad, V. S. L'vov, E. Podivilov and I. Procaccia, Phys. Rev. E {\bf 62}, 4901 (2000).

\bibitem{01AP}
I.Arad and I. Procaccia, Phys. Rev. E{\bf 63}, 056302 (2001).

\bibitem{01GHL}
S. Grossmann, A. Heydt and D. Lohse, J. Fluid. Mech, {\bf 440}, 381 (2001).

\bibitem{67Lum}
J. L. Lumley, Phys. Fluids {\bf 8}, 1056 (1967).


\bibitem{68Kra}
R.H. Kraichnan, Phys. Fluids {\bf 11}, 945 (1968).


\end{thebibliography}
\end{document}